\documentclass[prd,a4paper,nofootinbib,
showpacs, 
twocolumn,
]{revtex4}
\usepackage{graphicx}
\usepackage{amsfonts}
\usepackage{amssymb}
\def\eq#1{{eq.~(\ref{#1})}}

\def\etc{{\it etc.}}
\def\etal{{\it et al.}}
\def\eg{{\it e.g.}}

\newcommand{\be}{\begin{equation}}
\newcommand{\ee}{\end{equation}}
\newcommand{\bea}{\begin{eqnarray}}
\newcommand{\eea}{\end{eqnarray}}
\newcommand{\nn}{\nonumber}

\begin{document}
\title[]{Low-scale seesaw and  dark matter
}
\date{\today}
\author{M. Fabbrichesi$^{\ddag}$}
\author{S. Petcov$^{\dag\ddag\circ}$}
\affiliation{$^{\ddag}$INFN, Sezione di Trieste}
\affiliation{$^{\dag}$SISSA, via Bonomea 265, 34136 Trieste, Italy}
\affiliation{$^{\circ}$ Kavli IPMU, University of Tokyo, Japan}
\begin{abstract}
\noindent   We discuss how two birds---the little hierarchy problem of   low-scale type-I seesaw models and the search for a viable dark matter candidate---are (proverbially) killed by one  stone: a new inert scalar state
\end{abstract}

\maketitle
Together with the presence of dark mater (DM), neutrino oscillations---and the small neutrino mass  entailed by them---are the only physics beyond the standard model (SM)  experimentally confirmed. 

The most attractive model to account for the smallness of the neutrino masses is the seesaw mechanism~\cite{seesaw}. This mechanism requires  right-handed (RH) neutrinos  whose masses  can be taken at the GUT scale or, in low-scale scenarios, at lower energies if the Yukawa couplings are taken proportionally smaller, for instance, of the order of those of the charged leptons.

The inclusion of these new states within the SM  induces a finite renormalization that  tends to pull  the Higgs boson mass (or, equivalently,  the electroweak (EW) scale)  toward  the higher scale. This is not a problem if the new states are themselves at the EW scale and the renormalization is itself of the order of the Higgs boson mass. If instead the new states are at a larger scale, they give rise to a hierarchy problem in which to keep the  Higgs boson mass to its values we have to cancel the renormalization to the higher scale.

Rather than accomplishing such a cancellation  by a simple redefinition of the bare masses---a procedure that would require a cancellation between UV and IR degrees of freedom which seems artificial  even though technically possible---it is more appealing and practical to use the hierarchy problem in a heuristic manner to help us  in the definition of whatever model of physics we assume to exist beyond the SM~\cite{bf}.

In this Letter we discuss the case in which the masses of the RH neutrinos are in the range between 1 and 10 TeV. The renormalization effects are in this case dominated by the one-loop order and we  have what has been called the little hierarchy problem. To avoid this problem,  new states in addition to the RH neutrinos  must be included. As a matter of fact, the addition of one new state is enough. We study its properties and to what extent is a viable candidates for DM. This way, two birds (the little hierarchy problem and DM) are killed by one stone (the new state).

To avoid confusion, let us stress that  the hierarchy problem is often discussed in terms of the quadratic divergence arising in the mass term of the Higgs boson in a momentum dependent regularization (or, equivalently, in a pole in $d=2$ dimensions in dimensional regularization). The presence of these divergences makes the Higgs boson mass extremely sensitive to the UV physics and some cancellation must take place either in a natural manner by assuming a symmetry (usually, supersymmetry) or by fine-tuning by  imposing the Veltman condition~\cite{veltman}---namely that the new sector couples to the SM Higgs boson just so as to make  the  quadratic divergences to the SM Higgs boson mass vanish (see \cite{kundu} for various applications of this idea).  This is not the hierarchy problem we discuss in this Letter.  

The point of view we follow  is that all quadratic divergencies are a scheme-dependent artifact (similar to the  quadratic divergence arising in QED when we take a momentum dependent regularization  which  violates  gauge symmetry). In the Higgs boson mass case, they arise because of  the explicit breaking of scale invariance in momentum dependent regularizations, and should be eliminated by an appropriated counterterm~\cite{bardeen} or by not using that particular regularization scheme.
The point is that, even without these divergent terms, there are large finite renormalization effects which only depends on  integrating out the heavy modes in the low-energy effective theory---the SM in our case.  We identify the little hierarchy problem with the presence of these finite terms. These terms  are   similar to  those arising in a supersymmetric theory with soft mass terms where the quadratic divergencies are  cancelled while, after integrating out the heavy states, there are finite terms   whose contribution shifts the values of the Higgs boson mass. This Letter is about  these terms in the  case of the seesaw mechanism.


We consider a type I seesaw model in which three RH neutrinos
$N_{aR}$ are added to the SM as $SU(2)_L$ singlets.
The  lagrangian of the model  is given by the kinetic and Yukawa terms of the SM with  the addition of the neutrino Yukawa terms:
\be
\label{L}
\mathcal{L}=\,-\,  y_{a\ell}^\nu \bar{N}_{aR} \tilde{H}^\dag L_\ell
- \frac{1}{2} \bar N^c_{aL} M_{N ab}  N_{bR} + H.c.
\,,
\ee
where  $L_\ell$ represents the SM
left-handed $SU(2)$ doublet
$(\nu_\ell, \ell)_L$ and $\ell = e, \mu, \tau$.
In \eq{L}, the Yukawa term gives rise to the
neutrino Dirac mass matrix, $M_D = y\,v_W$,
 after the Higgs field $H=(v_W+h)$ takes its vacuum expectation value $v_W=174$ GeV.
The heavy RH neutrinos $N_{aR}$ have a Majorana mass term.

  We compute the one-loop  finite correction to the Higgs boson
mass using dimensional regularization with renormalization scale $\mu$.
The SM particle contributions are negligible. To compute the one-loop
renormalization arising from the heavy Majorana  neutrinos,
we rotate the Yukawa couplings $y^\nu_{al}$ into the basis in which
 the heavy RH neutrino mass matrix $M_N$ is diagonal.
In this basis the matrix of neutrino Yukawa couplings takes the following
form in the type I seesaw model of interest \cite{Ibarra:2010xw,Cely:2012bz}:
\be
\hat{y}^\nu_{j\ell}= M_{N_j}(RV)^T_{j\ell}/v_W\,,
\ee
 where $V$ is a unitary matrix which diagonalises the
RH neutrino Majorana mass matrix,
$M_N = V \hat{M} V^T$ with $\hat{M} = \mbox{diag} (M_1,M_2,M_3)$,
$M_j$ being the mass of the heavy neutrino mass-eigenstate $N_j$,
and $R^T \cong M^{-1}_N\,M_D$ ($|M_D| \ll |M_N|$).
As can be shown (see, \eg, \cite{Ibarra:2010xw}),
the quantity $(RV)_{\ell j}$
represents the weak charged current and neutral current
coupling of the heavy Majorana neutrino
$N_j$ to the charged lepton $l$ and the $W^{\pm}$-bosons,
and to the LH flavor neutrino $\nu_l$ and the $Z^0$-boson.
The matrix $\eta = -\,0.5 (RV)(RV)^\dagger$ describes, in
the seesaw model considered,
the deviations from unitarity of the Pontecorvo, Maki,
Nakagawa, Sakata (PMNS) neutrino mixing matrix $U_{\rm PMNS}$:
$U_{\rm PMNS} = (1 + \eta)U$, where $U$ is a unitary matrix which
diagonalises the Majorana mass matrix of the LH flavor neutrinos,
 $m_{\nu}$, generated by the seesaw mechanism.

In the type I seesaw scenario, the elements of
the matrix $(RV)$ are bounded by their relation to
the elements of the neutrino mass matrix $m_{\nu}$ \cite{Ibarra:2010xw},
which all have to be smaller than approximately 1 eV:
\be
\label{ev}
\left| \sum_k (RV)^*_{\ell'k} M_k (RV)^\dag_{k\ell} \right| = |(m_v)_{\ell'\ell}|
\lesssim 1 \; \mbox{eV}\, .
\ee

In the traditional seesaw model the Yukawa
couplings are taken typically
to be of order one and the masses $M_{N_j}$ are
very large, close to the GUT scale.
The couplings  $|(RV)_{\ell k}|$ in
this case have to be very small to satisfy \eq{ev}.
In low-scale seesaw models, the heavy Majorana neutrino masses lie at the TeV scale
and the couplings $|(RV)_{\ell k}|$ are proportionally larger.
In this scenario, $|(RV)_{\ell k}|$ can even be larger if there is partial or complete
cancellation between the terms in the sum in the r.h.s. of \eq{ev}.
This possibility can be realized, \eg, in models \cite{Ibarra:2011xn}
with two heavy Majorana neutrinos $N_1$ and $N_2$,
which have relatively close masses,
$M_2 = M_1 (1 + z)$, $z \ll 1$, thus forming a pseudo-Dirac state \cite{LWPD81},
and whose couplings $(RV)_{\ell 1}$ and $(RV)_{\ell 2}$ are related:
$(RV)_{\ell 2}\sqrt{M_2} = \pm i\,(RV)_{\ell 1}\sqrt{M_1}$, $l=e,\mu,\tau$.
The indicated conditions can take place, for instance, in theories with
an approximately conserved lepton charge (see, \eg, \cite{Gavela:2009cd}).

In the scenario with two heavy Majorana
neutrinos outlined above, the flavor structure of the couplings
$(RV)_{\ell j}$, $j=1,2$, is completely determined by the
requirement of reproducing the neutrino oscillation data
and the scheme is characterized by four parameters \cite{Ibarra:2011xn}:
$M_1$, $z$, the largest eigenvalue $y$
of the matrix of neutrino Yukawa couplings (see further) and
a CP violation phase. The neutrino oscillation data,
the EW precision measurements and the existing limits
on the rates of lepton
flavor violating (LFV) processes
involving the charged leptons (as  the
$\mu \rightarrow e + \gamma$, $\mu \rightarrow 3e$ decays, \etc),
imply the following upper bounds on
the couplings $|(RV)_{\ell 1}| \cong |(RV)_{\ell 2}|$
(see, \eg, \cite{Dinh:2012bp,Akhmedov:2013hec}
and references quoted therein):
\be
\label{exp1}
|(RV)_{e1}|^2\,,|(RV)_{\mu1}|^2\,,|(RV)_{\tau1}|^2 \lesssim 10^{-3} \,
\ee
where we have quoted a somewhat simplified
constraint on the three couplings.
The actual upper bounds depend 
on the flavor index $l$ of the couplings,
but the variation with $l$ is not significant and for the purposes
of our investigation it can be neglected. We will use the
generic bounds given in \eq{exp1} in our analysis.

  In what follows we will neglect for simplicity the splitting
between the two heavy Majorana neutrino masses $z$, i.e.,
we will set $z = 0$ and will use $M_1 = M_2 \equiv M_N$.
The corrections due to $z\neq 0$ are insignificant
in the problem of interest.
For $z=0$, the largest eigenvalue $y$
of the matrix of neutrino Yukawa couplings
is given by \cite{Ibarra:2011xn}
\be
\label{largest}
y^2 v_W^2 = 2 M_N^2 \left[ |(RV)_{e1}|^2 + |(RV)_{\mu1}|^2
+ |(RV)_{\tau1}|^2\right]
\ee

Taking into account the one-loop contribution, and assuming
RH neutrino degeneracy, the Higgs boson mass receives a shift given by
\be
\label{finite}
 \delta\mu_H^2 (\mu) = \frac{4 y^2}{ (4 \pi)^2}  M^2_{N} \left(1 - \log\frac{M^2_{N}}{\mu^2}\right) \, ,
\ee
being $\mu$ the matching scale that in this case we can identify with $M_N$.

The addition of the RH neutrinos would shift the Higgs boson mass to the new scale unless we  balance this new contribution to prevent large one-loop renormalizations.  The identification of what states (their  masses and couplings to the Higgs boson)  must be present for such a balancing act to occur  provides the heuristic power of the little hierarchy problem.

While many possible new states can be added to prevent large corrections to the Higgs boson mass, the simplest choice consists in including just an  inert scalar state~\cite{inerts,Yaguna:2008hd}, that is, a scalar particle only interacting with the Higgs boson (and gravity)---and therefore  transforming as the  singlet representation of the EW gauge  group $SU(2)\times U(1)$ (and similarly not charged under the color group)----which acquire no vacuum expectation value. Such a choice minimizes unwanted effects on EW radiative corrections and other physics well described by the SM.

If in addition we impose a $Z_2$ symmetry under which  the  inert scalar is odd and all the SM fields are even,  the new state will couple  to the SM Higgs doublet only through quartic interactions in the scalar potential. By construction, as only look for solutions with  vanishing vacuum expectation value, the symmetry $Z_2$ remains unbroken and after EW symmetry breaking the singlet state can, as we shall discuss, potentially be a viable  cold DM candidate.

The scalar potential of the model is given by
\bea
\label{pot}
V(H,S)&=&\mu_H^2 (H^\dag H)+ \mu^2_S S^2\nn\\
& +&\lambda_1 (H^\dag H)^2 +
\lambda_{2} S^4+ \lambda_{3} (H^\dag H ) S S \,.
\eea
Linear and trilinear terms for $S$ are absent due to the $Z_2$ symmetry mentioned above.

Taking into account the one-loop contribution induced by the  scalar state $S$,  the overall shift to $\mu^2_H$, taking $M_S<M_N$ and $\mu=M_S$ to minimize the logarithmic contributions to the matching,  becomes
\bea
\label{finite2}
 \delta\mu_H^2(M_S) &=& \frac{1}{(4 \pi)^2}\Big[ \lambda_3 M_S^2 \nn  \Big.\\
&-&  \left.  4 y^2 M_N^2  \left(1 - \log\frac{M^2_{N}}{M^2_S}\right)\right]\,,
\eea
where we have taken $y$ given in \eq{largest} for the Yukawa couplings.

We want the  correction in \eq{finite2} to be of the order of the Higgs boson mass itself. For simplicity, we can just impose that $\delta \mu^2_H=0$ and obtain
\be
\label{l3}
\lambda_3=\frac{4 y^2 M_{N}^2}{M_S^2} \left(1 - \log \frac{M_N^2}{M_S^2} \right)\, .
\ee
Equation (\ref{l3}) is economical but we must bear in mind that it represents  just a special case in which  the one-loop renormalization exactly vanishes. More solutions can easily be found for $\delta \mu^2_H \simeq m_h^2$ but do not change  in a significant manner the numerical results.

Because of the extra factor $1/(4\pi)^2$, two-loop corrections become important only if the masses are above 10 TeV. To be safe, we take the matching scale $\mu=M_S$ smaller than 7 TeV.


\begin{figure}[ht!]
\begin{center}
\includegraphics[width=2.5in]{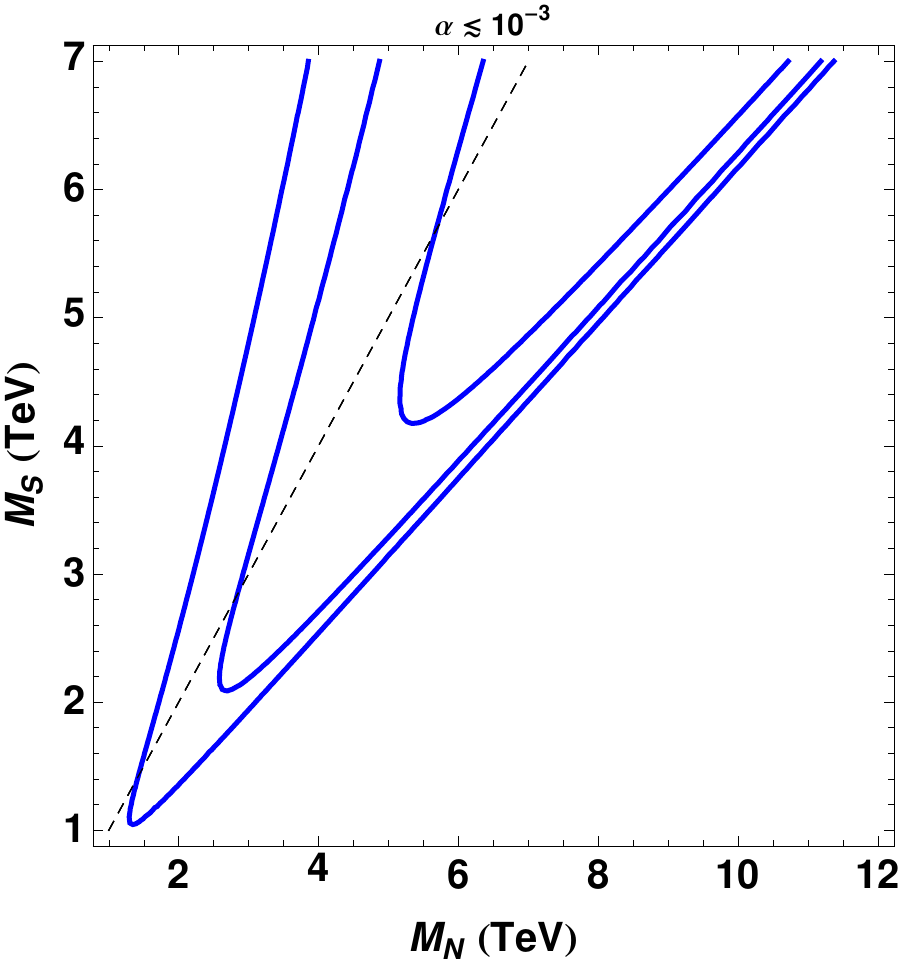}
\includegraphics[width=2.5in]{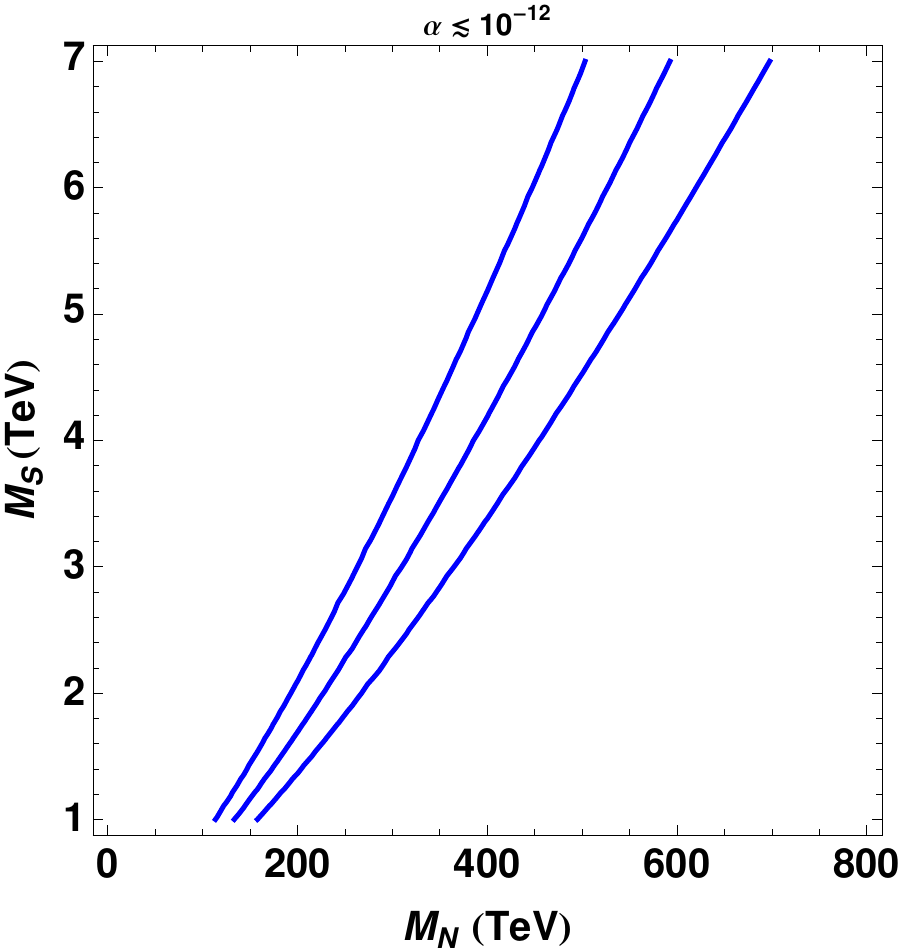}
\caption{\small  Contour plot for the WIMP solution: Values of $M_S$ and $M_N$ that are solutions of \eq{sol} for  two limiting sets of  values of the parameter $\alpha$. On top, (in descending order) $\alpha = \{1, 2, 4\} \times  10^{-4}$,  below $\alpha = \{1, 2, 4\} \times 10^{-13}$. Solutions with $m_S < m_N$ are below the dashed line in the figure on top.
\label{figA}}
\end{center}
\end{figure}

It has been shown~\cite{Yaguna:2008hd} that a single inert singlet  that couples with the Higgs boson with a small coupling    is a realistic cold weakly interacting DM candidate (WIMP) with  a mass below $v_W$. In our case,  the singlet  may account for the correct relic density in the opposite regime where  its mass is much larger than $v_W$ and its coupling with the Higgs boson   relatively large.

In this case,  the  scattering amplitude is dominated by the point-like  $S S \to hh$ vertex  which gives a contribution to the thermally averaged total cross section equal to
\be
\label{sigma}
\langle \sigma v \rangle  \simeq \frac{1}{4 \pi} \frac{\lambda_{3 }^2}{M_{S }^2}\, \sqrt{1 - \frac{m_h^2}{m_S^2}} \,,
\ee
where we keep only the s-wave contribution.

To estimate the viability of $S $ as DM candidate,  we must compute its relative relic abundance~\cite{density}
\be
\Omega_S = \frac{M_S n_S(t_0)}{\rho_c}
\ee
where $\rho_c = 1.05 \, h^2\,  10^{-5}$ GeV/cm$^3$ and the density $n_S(t_0)$ is  given by
\be
n_S(t_0) = \sqrt{\frac{45}{\pi g_{*}}}\frac{s_0}{M_{pl} T_f \langle \sigma  v\rangle}\,,
\ee
where $M_{pl}$ is the Planck mass, $T_f$ is the freeze-out  temperature, which for our and similar  candidates  is approximately given by
\be
m_{S }/T_f \simeq \log \frac{M_S M_{pl}\langle \sigma v \rangle  }{240 \sqrt{g_*}} \sim 26 \,,
\ee
and $s_0 = 2.8 \times 10^{3}$ cm$^{-3}$ is the entropy density. The constant $g_*= 106.75+1$  counts the number of  SM degrees of freedom in thermal equilibrium plus the additional degrees of freedom related to the singlet.

We therefore obtain
\be
\label{value}
\Omega_S h^2 \simeq  8.41 \times 10^{-11} \frac{M_{S }}{T_f } \sqrt{\frac{45}{\pi g_{*}}} \frac{\rm{GeV}^{-2}}{\langle \sigma  v\rangle}  \, ,
\ee
which is sufficiently accurate for our purposes.

Current data fit within the standard cosmological model  give a relic abundance with $\Omega_{\mathrm{DM}} h^2=0.1187\pm 0.0017$~\cite{planck}.
By combining  the central value above with \eq{sigma} and \eq{value}, we can write the coupling $\lambda_3$ as function of $M_S$ thus obtaining
\be
\label{lambda3}
|\lambda_3|\simeq 0.15 \, \frac{M_S}{\mbox{TeV}}\,.
\ee
This solution gives a  DM candidate which can account for 100\% of the relic density and with a cross section of a few pb which makes it  weakly interacting, a WIMP.

 The presence of the scalar singlet DM improves  the EW vacuum stability with respect to the SM~\cite{Gonderinger:2009jp}. This is particularly interesting in connection with the presence of the RH neutrinos which, in general, have the opposite effect of reducing the stability region of the Higgs boson potential~\cite{EliasMiro:2011aa}.

\begin{figure}[t!]
\begin{center}
\includegraphics[width=2.5in]{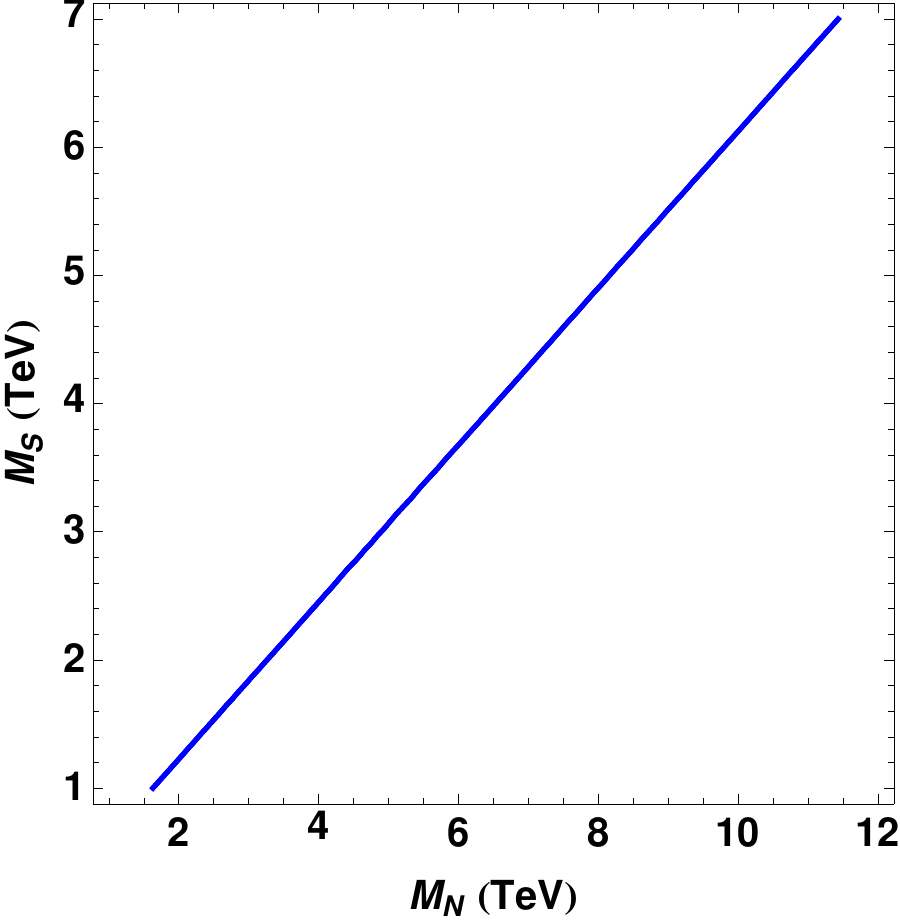}
\caption{\small   Contour plot for the FIMP solution: Values of $M_S$ and $M_N$ that are solutions of \eq{l3} in the case of $\lambda_3 \simeq 10^{-11}$ and $\alpha \simeq 10^{-12}$.
\label{figB}}
\end{center}
\end{figure}

Insertion of \eq{lambda3} in \eq{l3} gives a relationship between the RH neutrino mass and that of the new scalar if the latter is to be considered a viable candidate for DM:
\be
\label{sol}
0.15\, M_S^3 = 8 \,\alpha \,\frac{M_N^4}{v_W^2}   \left(1 - \log\frac{M^2_{N}}{M_S^2}\right)
\ee
where $\alpha =|(RV)_{e1}|^2 + |(RV)_{\mu1}|^2 + |(RV)_{\tau1}|^2$ represents the sum of the squares of the couplings of the RH and LH neutrinos. In \eq{sol}, all masses must be taken in TeVs.

Possible solutions are shown in  Fig.~\ref{figA}. We can see that we can have solutions with $M_S \simeq M_N$  as long as we take the couplings  between LH and RH neutrinos, and therefore $\alpha$, to be as large as possible within the experimental constraints in \eq{exp1}. This case corresponds to taking the largest Yukawa coupling just at its experimental bound. This is the most interesting range because the RH neutrino masses are still in a range accessible to the experiments (\eg, $\mu \rightarrow e + \gamma$, $\mu \rightarrow 3e$ decays, 
$\mu^- - e^-$ conversion in nuclei, neutrinoless double $\beta$-decay).

On the other hand, if these couplings are taken at their natural values (and no cancellation is assumed in their sum) we can have only solutions where $M_S$ is much smaller than $M_N$ because now the Yukawa couplings are much smaller than 1.   In this case,  $\lambda_3$ becomes negative and  one has to check that $\lambda_3 \geq -2 \sqrt{\lambda_1 \lambda_2}$ for the stability of the potential. This  scenario seems less interesting than the previous  one  because
the RH neutrino masses are  several hundreds of TeVs and therefore outside the range of any foreseeable experiment.

  It is interesting to notice that there exists another regime in which the inert singlet is a viable DM candidate. When  $\lambda_3$ is much smaller than 1, thermal equilibrium for the scalar states is never attained and their abundance is so low that they never annihilate among themselves. The usual result does not apply. In this case, the DM candidate is what has been called a feebly interacting massive particle (FIMP).
 If we do not use \eq{lambda3} and take $\lambda_3 \simeq 10^{-11}$~\cite{fimp},  we can have a DM candidate in which $M_S \simeq M_N$ (see Fig.~\ref{figB}). For these solutions, $\alpha \simeq 10^{-12}$ and  the Yukawa couplings become of the order of those of the charged leptons.

Let us   briefly  comment on the possibility
 of detecting the inert scalar $S$ in the case in which is a WIMP. The multi TeV range  of its mass makes its detection at the LHC very difficult if not impossible. Whereas a detailed discussion of its possible role in the phenomena observed
by the current experiments in space (PAMELA, FERMI, AMS2 \etc) is beyond the scope of this Letter,  nuclear scattering experiments are more promising and easier to quantify. 
  The  quartic term proportional to $\lambda_3$  in \eq{pot} gives rise, after EW symmetry breaking,  to the three-field interaction $SSh$ which yields the effective singlet-nucleon vertex
$
f_N  \lambda_3 m_N/m_h^2 \, S S \,\bar{\psi}_N \psi_N\,,
$
where $m_N$ is the nucleon mass and the  factor $f_N$ contains many uncertainties due to the computation of the nuclear matrix elements and it can vary  from  0.3 to 0.6~\cite{nucleon}.
The (non-relativistic) cross section for the process  is given by~\cite{sigma}
\be
\sigma_N = f_N^2 m_N^2 \frac{\lambda_3^2}{4 \pi} \left( \frac{m_r}{m_{S} m^2_h} \right)^2 \,,
\ee
where $m_r$ is the reduced mass for the system which is, to a vary good approximation  in our case, equal to the nucleon mass $m_N$. Substituting the values we have found for our model and  for $m_S$ of a few  TeV, we obtain, depending on the choice of parameters and within the given uncertainties, a cross section
$\sigma_N$  of order  $10^{-45}\mbox{cm}^2$, a value  within reach of the next generation of experiments~\cite{DMT}.

\acknowledgements

This research was supported in part by the World Premier
International Research Center Initiative (WPI Initiative),
MEXT, Japan, by the European Union FP7-ITN INVISIBLES
(Marie Curie Action, PITAN-GA-2011-289442) and
by the INFN program on ``Astroparticle Physics'' (S.T.P.).



\end{document}